Systematics of the strength of superconductivity of the *d-p* model in relation to cuprates

K. Yamaji* and T. Yanagisawa

*Electronics and Photonics Research Institute, AIST, 1-1-1 Umezono, Tsukuba 305-8568, Japan*

Abstract: Cuprate high-$T_c$ superconductors are known to have several relationships between $T_c$ and features of electronic structure. In the first found one, $T_c$ is correlated with the Madelung potential difference between planar Cu and O sites. Another one relates $T_c$ to the relative ratio of hole densities at the Cu and O sites. In this work we analyze the dependences of the strength of superconductivity (SC) on the model parameters of the *d-p* model. We first transform the *d-p* model into a Hubbard model which proves to have a ***k***-dependent Coulomb interaction. To this model we apply a kind of mean field theory, i. e., Kondo's theory of the *d*-wave SC gap in cuprates. The obtained so-call SC strength parameter increases with increase of $|\varepsilon_p-\varepsilon_d|$ on both positive and negative sides of $\varepsilon_p-\varepsilon_d$ due to the increase of effective on-site Coulomb interaction $U$; $\varepsilon_p$ and $\varepsilon_d$ are hole energy levels of O2*p* and Cu3*d* orbital's referenced to the $d^9$ configuration. The above-mentioned two relationships are shown to be undertandable in terms of the *d*-wave SC if $\varepsilon_p-\varepsilon_d<0$.

*Corresponding author:
Name: Kunihiko Yamaji
Email address: yamaji-kuni@aist.go.jp

# 1. Introduction

Relationships between $T_c$ and features of electronic structure in cuprate high-$T_c$ superconductors are fascinating. Torrance and Metzger [1] found the first such relationship between $T_c$ and the Madelung potential difference, $\Delta V_M$, between the Cu and O sites in the CuO$_2$ plane. $T_c$ was found to increase with decreasing $\Delta V_M$. There is another correlation which reveals increasing $T_c$ with increasing relative ratio of hole density at O site against that at Cu site [2]. Among several known relationships this work aims to show that these two relationships are understandable on the basis of the *d-p* model. This model includes the O$2p_\sigma$ orbital explicitly as an additional electronic freedom compared with the Hubbard model. Further, it is known that the effective on-site Coulomb energy $U$ derived from such multi-orbital model considerably depends on the band and atomic parameters [3]. $U$ can change appreciably, depending on the band situation. We started to try to explain the main features of the strength of superconductivity (SC) from the dependence of effective $U$ on the band situation.

In the first step, we transform the *d-p* model into a one-band model, i.e., a kind of Hubbard model. It is found to have an interesting ***k***-dependent Coulomb interaction. Then, we compute a SC strength parameter $x$, which we call so, for this model with the aid of Kondo's gap equation for the SC gap parameter [4,5], which is a basically a mean field theory. We employ it in a wide range of $\varepsilon_p - \varepsilon_d$; $\varepsilon_p$ and $\varepsilon_d$ are energy levels of O $2p_\sigma$ and Cu $3d_{x^2-y^2}$ orbitals, respectively, in the hole picture. Although Kondo's gap equation was proved in the weak-coupling case, we presume that results extrapolated to intermediate-coupling situation allow us to qualitatively compare relative SC strengths among superconductors having different $\varepsilon_p - \varepsilon_d$ values. We treat the *d*-wave SC. The obtained SC strength parameter for the b$_{1g}$-type increases with increase of $|\varepsilon_p - \varepsilon_d|$ on both positive and negative sides of $\varepsilon_p - \varepsilon_d$ due to the increase of effective $U$. In fact when $\varepsilon_p - \varepsilon_d < 0$, the b$_{2g}$-type SC in stead of b$_{1g}$ take a slightly larger $x$ value with our model parameters, although the $\varepsilon_p - \varepsilon_d$ dependences of the both are similar. From the dependence of the SC strength on $\varepsilon_p - \varepsilon_d$, the above-mentioned two systematics are understandable if we choose parameter values satisfying $\varepsilon_p - \varepsilon_d < 0$. The present approach sheds light not only to the two correlations but also a new aspect of the mechanism of high-$T_c$ of cuprates.

The initiating work of the present paper was reported in a preceding paper [6]. In the present paper equations and computations are developed and described more extensively. Computations with more transparent parameter sets are added so that the results are more persuasive.

## 2. *d-p* model and derivation of *k*-dependent Hubbard model

Our *d-p* model is the conventional one based on the hole picture [7]. It gives the following one-particle part:

$$H_0 = \sum_{k\sigma} \left(d_{k\sigma}^\dagger, \; p_{xk\sigma}^\dagger, \; p_{yk\sigma}^\dagger\right) \begin{bmatrix} \varepsilon_{dk} & t_{dp}[1-\exp(ik_x)] & t_{dp}[\exp(ik_y)-1] \\ t_{dp}[1-\exp(-ik_x)] & \varepsilon_{pk} & \\ t_{dp}[\exp(-ik_y)-1] & & \varepsilon_{pk} \end{bmatrix} \begin{bmatrix} d_{k\sigma} \\ p_{xk\sigma} \\ p_{yk\sigma} \end{bmatrix}. \quad (1)$$

The 3×3 matrix is denoted as $\hat{H}_0(k)$. $d_{k\sigma}^\dagger$ ($d_{k\sigma}$) is the creation (annihilation) operator for the *d*-hole. They are Fourier-transformed from the real-space operator $d_{R\sigma}^\dagger$ ($d_{R\sigma}$) for site *R* on the two-dimensional (2D) square lattice. $p_{xk\sigma}$ and $p_{yk\sigma}$ are operators Fourier-transformed for $p_{x,R+\hat{x}/2,\sigma}$ and $p_{y,R+\hat{y}/2,\sigma}$, respectively; $\hat{x}$ and $\hat{y}$ denote the unit lattice vectors along the *x*- and *y*-directions, respectively. $\varepsilon_{pk}$ and $\varepsilon_{dk}$ are the energy levels for the O *p*- and Cu *d*-orbitals, respectively, referenced to the $d^9$ configuration. They are shifted due to inter-unit-cell hoppings from unmodified levels $\varepsilon_p$ and $\varepsilon_d$ in the following way:

$$\begin{aligned}\varepsilon_{pk} &= \varepsilon_p - 4\tau' \cos k_x \cos k_y - 2\tau_2\left(\cos(2k_x) + \cos(2k_y)\right), \\ \varepsilon_{dk} &= \varepsilon_d - 4\tilde{t}' \cos k_x \cos k_y - 2\tilde{t}_2\left(\cos(2k_x) + \cos(2k_y)\right).\end{aligned} \quad (2)$$

Hamiltonian $H_{\text{int}}$ for the Coulomb interaction consists of three parts with on-site Coulomb energies $U_d$, $U_p$ and $U_p$, respectively.

The eigenvalues of $\hat{H}_0(k)$ is given by

$$\lambda_0 = \varepsilon_{pk} \quad \text{and} \quad \lambda_\pm = E_\pm(k) = \tfrac{1}{2} D_k + \varepsilon_{dk} \pm S_k, \quad (3)$$

where $D_k \equiv \varepsilon_{pk} - \varepsilon_{dk}$ and $S_k = \left\{D_k^2/4 + 4t_{dp}^2\left[\sin^2(k_x/2) + \sin^2(k_y/2)\right]\right\}^{1/2}$. The lowest eigenvalue is $E_-(k)$ and makes the band accommodating holes. Two higher bands originates from $\lambda_0$ and $\lambda_+$. $E_-(k)$ is approximately given by

$$E_-(k) \cong \text{const} + 2t(\cos k_x + \cos(k_y)) - 4t' \cos k_x \cos k_y - 2t_2\left(\cos(2k_x) + \cos(2k_y)\right), \quad (4)$$

where const=$\varepsilon_d-4t$, $t = t_{dp}^2/|\varepsilon_p - \varepsilon_d|$, $t' = \tilde{t}'$ and $t_2 = \tilde{t}_2$ in the case of $\varepsilon_p - \varepsilon_d > 0$; in the case of $\varepsilon_p - \varepsilon_d < 0$, const=$\varepsilon_p - 4t$, $t = t_{dp}^2/|\varepsilon_p - \varepsilon_d|$, $t' = \tau'$ and $t_2 = \tau_2$. Here we retain the terms down to the first order in $1/(\varepsilon_p - \varepsilon_d)$, $t'$ or $t_2$. Components of the corresponding eigen vector ${}^t(u_-(k), v_-(k), w_-(k))$ for $\hat{H}_0(k)$ are given as

$$u_-(k) = \left[(1 + D_k/2S_k)/2\right]^{1/2}, \tag{5a}$$

$$v_-(k) = \left[(1 - D_k/2S_k)/8\right]^{1/2} \left[\exp(-ik_x) - 1\right] / \left[\sin^2(k_x/2) + \sin^2(k_y/2)\right]^{1/2}, \tag{5b}$$

$$w_-(k) = \left[(1 - D_k/2S_k)/8\right]^{1/2} \left[1 - \exp(-ik_y)\right] / \left[\sin^2(k_x/2) + \sin^2(k_y/2)\right]^{1/2}. \tag{5c}$$

Similarly we get eigenvectors ${}^t(u_+(k), v_+(k), w_+(k))$ and ${}^t(u_0(k), v_0(k), w_0(k))$ corresponding to $\lambda_+ = E_+(k)$ and $\lambda_0 = \varepsilon_{pk}$, respectively. They compose the unitary matrix which transforms $d_{k\sigma}$, $p_{xk\sigma}$ and $p_{yk\sigma}$ into $\alpha_{-,k\sigma}$, $\alpha_{+,k\sigma}$, and $\alpha_{0,k\sigma}$, diagonalizing $\hat{H}_0(k)$.

In the process of transformation of the *d-p* model to a Hubbard model, we retain only terms containing the $\alpha_{-,k\sigma}$ operator of the lowest band, neglecting other terms. Then, we get $d_{k\sigma} \approx u_-(k)\alpha_{-,k\sigma}$, $p_{xk\sigma} \approx v_-(k)\alpha_{-,k\sigma}$ and $p_{yk\sigma} \approx w_-(k)\alpha_{-,k\sigma}$. This is justified since both intrinsic and doped holes occupy the lowest band almost all the time and important hole dynamics is given by this $\alpha_{-,k\sigma}$ operator part of the Hamiltonian. There are 1.15 holes per unit cell. In the case of $\varepsilon_p - \varepsilon_d > 0$, where holes occupy mainly the Cu *d* orbitals, coefficients $v_-(k)$ and $w_-(k)$ decrease very quickly with increase of $\varepsilon_p - \varepsilon_d$. Therefore, we neglect *p* operators and so $U_p$ as well. Then, $H_{int}$ is reduced to only one term proportional to $U_d$ as follows:

$$H_{int} = (1/N) \sum_{k_1 k_2 k_3 k_4} \delta(k_1 + k_3, k_2 + k_4) U_d u_-^*(k_1) u_-(k_2) u_-^*(k_3) u_-(k_4) \alpha_{-,k_1\uparrow}^\dagger \alpha_{-,k_2\uparrow} \alpha_{-,k_3\downarrow}^\dagger \alpha_{-,k_4\downarrow}. \tag{6}$$

Thus, our *d-p* model is transformed to a Hubbard model with adjustable band parameters but with a *k*-dependent Coulomb interaction. The interaction term depends on four *k*-arguments, which is found to make the interaction considerably stronger in the scattering process around the node and weaker around the antinode. In the case of $\varepsilon_p-\varepsilon_d>0$ and $t'\sim 0.5$, it was found to make the scattering process in the nodal region stronger by about 20 % and weaker by the same amount in the antinodal region.

## 3. Kondo's gap equation and its modification for our model in the case of $\varepsilon_p-\varepsilon_d>0$

Kondo demonstrated that the ground state of the standard two-dimensional Hubbard model is a BCS-type superconducting state with $d$-wave pairing in the small $U$ range if the electron density lies between 0.6 and 0.9 per site [4,5]. This occurs since a pairing interaction expressed as

$$V_{kk'} = (U/N) + (U^2/N)\chi(k+k') \qquad (7)$$

works in the SC gap equation

$$\Delta_k = -\sum_{k'} V_{kk'} \Delta_{k'}/2E_{k'}, \qquad (8)$$

where $\chi$ is the wave-number dependent spin susceptibility and $E_k=[(\varepsilon_k-\mu)^2+|\Delta_k|^2]^{1/2}$; $\mu$ is the chemical potential. Diagrammatic processes generating this pairing interaction are shown in Fig. 1. $k$ and $-k$ are wave vectors denoting a pair of incoming electrons with opposite spins. $k'$ and $-k'$ denote wave vectors of a pair of outgoing electrons. $N$ is the number of electron sites on the square lattice. In the present model with the modified Coulomb interaction, the same scheme works if we replace $U$ appropriately, e. g., by $U_d u_-(k) u_-(-k) u_-^*(k') u_-^*(-k')$. $V_{kk'}$ is replaced by

$$V_{kk'} = (U_d/N)\left(1+U_d\chi_{uu}(k+k')\right)|u_-(k)|^2|u_-(k')|^2, \qquad (9)$$

$$\chi_{uu}(q) \equiv \frac{2}{N}\sum_k \frac{f_k(1-f_{k+q})}{E_-(k+q)-E_-(k)} |u_-(k)|^2|u_-(k+q)|^2$$

$$= \frac{1}{N}\sum_k \frac{f_k - f_{k+q}}{E_-(k+q)-E_-(k)} |u_-(k)|^2|u_-(k+q)|^2. \qquad (10)$$

$\chi_{uu}(q)$ is defined by eq. (10). Fermi-Dirac distribution function $f_k$ is for the quasi-particle with $E_-(k)$ at temperature $T$. The first order term in $U_d$ can be neglected since they give no contribution in the $d$-wave SC.

The gap equation is an integral equation for solving $\Delta_k$ defined by eq. (8) in two-dimensional $k$ space. In the case of small $\Delta$, Kondo reduced it to a one-dimensional one with integration along the Fermi surface as follows:

$$z_k = \log\Delta \cdot \sum_{k'} V_{kk'}\delta(\varepsilon_{k'})z_{k'}, \qquad (11)$$

where gap parameter $\Delta_k$ is rewritten as $\Delta z_k$ with $z_k$ giving the $\theta$-dependence ($\theta$ is the in-plane angle of $\boldsymbol{k}$); $\varepsilon_k$ is the one-particle energy referred to the Fermi energy. The problem is reduced to a linear eigenvalue equation. This equation is Fourier-transformed and as its eigenvalue we get a SC strength parameter $x$ which is linked by definition to the amplitude $\Delta$ of $d$-wave gap parameter by

$$x = -2t^2/U_d^2 \log(\Delta). \qquad (12)$$

$x$ is a function increasing with increasing $\Delta$. We use eq. (12) with $U_d=8$ through Fig. 2 to 5 which show $x$ against $\varepsilon_p-\varepsilon_d$.

Below we mention several practical facts in carrying out the calculations. We employ hole band with a closed Fermi surface around $(\pi, \pi)$ instead of the electron band around $(0, 0)$; the lattice constant is the length unit. For numerical integration the Fermi surface was cut into 400~500 pieces. On-site Coulomb energy is fixed at $U_d=8$ and $U_p=6$ in the proper unit in each case. Since we looked for $\Delta$ at O K, temperature $T$ in the Fermi-Dirac function should be zero, but we employ $T=10^{-6}$ to avoid singularities in numerical integration. Up to 15$^{th}$ higher harmonics were used in the eigenvalue equation leading to the values of $x$ and $\Delta$ in eq. (12). The error bar coming from these numerical procedures is estimated less than 10 %

## 4. SC strength parameter $x$ in the case of $\varepsilon_p-\varepsilon_d>0$
### 4.1 First band parameter set

The result is shown in Fig. 2. The nearest-neighbor transfer energy $t = t_{dp}^2/|\varepsilon_p-\varepsilon_d|$ is fixed at unity so that $t$ plays the role of the energy unit. $t_{dp}$ changes together with $\varepsilon_p-\varepsilon_d$. $U_d$ is set at 8 by hand. For band parameters we take typical values for HgBa$_2$CuO$_4$ [8] and Tl$_2$Ba$_2$CuO$_6$ [9], i.e., $t'=0.25$ and $t_2=-0.125$ in order to reproduce the experimentally determined form of the Fermi surface. We assumed the SC of the type of even parity so that we assume the $d$-wave SC of the four types: a$_{1g}$, a$_{2g}$, b$_{1g}$ and b$_{2g}$. As seen in the figure, with increase of $\varepsilon_p-\varepsilon_d$ the SC strength $x$ increases very quickly. This means the quick increase of SC gap $\Delta$ and so of $T_c$. The $x$ value, therefore, $\Delta$ for b$_{1g}$ type increases overwhelmingly quickly in comparison with the a$_{1g}$, a$_{2g}$ and b$_{2g}$ types. Although the results are obtained with assumption of weak $U_d$, $x$ is extrapolated to the case of the intermediate $U_d$. This procedure is assumed to give the qualitatively correct relative strength of $x$ as a function of $\varepsilon_p-\varepsilon_d$.

4.2 Second band parameter set

We studied the case of another band parameter set in which $t_{dp}$ is set at 1; $t = t_{dp}^2/|\varepsilon_p - \varepsilon_d|$ is variable; $t'=0.25t$ and $t_2=-0.125t$. The ratio of $t$, $t'$ and $t_2$ is kept the same as in the case of the first band parameter set so that the form of the Fermi surface is approximately reproduced like that of $HgTl_2CuO_4$ and $Tl_2Ba_2CuO_6$. As seen in Fig. 3 the $x$ value increases significantly more quickly with increase of $\varepsilon_p - \varepsilon_d$. This setting allows more intuitive comparison of the results with experiments, since $t_{dp}$ is fixed. When we compare various cuprats, fixing $t_{dp}$ is more closer to experimental situation than fixing $t$. Therefore, the results in Fig. 3 confirm that $x$ clearly increases with increase of $\varepsilon_p - \varepsilon_d$ in the case of $\varepsilon_p - \varepsilon_d > 0$ in experimental situations on cuprates.

4.3 Comments

The tendency that x increases with increasing $\varepsilon_p - \varepsilon_d$ is understandable in terms of an effective on-site Coulomb interaction defined by

$$U_{\text{eff}} = U_d <|u_-(k)|^2>^2 \approx (U_d/4)\left\{1+1\left[1+16t_{dp}^2/(\varepsilon_p-\varepsilon_d)^2\right]^{1/2}\right\}^2. \quad (13)$$

If we replace $U_d|u_-(k)|^2|u_-(k+q)|^2$ in our version of Kondo's gap equation by $U_{\text{eff}}$, the obtained $x$ is very close to the above-mentioned result. This was confirmed in the first case. We expect the similar behavior in the second case. This clearly indicates that increase of $\varepsilon_p - \varepsilon_d$ leads to the increase of $U_{\text{eff}}$ and the latter increases $x$.

Such a positive correlation between SC strength (or gap parameter) and $\varepsilon_p - \varepsilon_d$ is consistent with the wide-range tendency in the result of the variational Monte Carlo calculation for the $d$-$p$ model by Koike et al. [10] and Yanagisawa et al. [7] in the case of $\varepsilon_p - \varepsilon_d > 0$.

We cannot change only $\varepsilon_p - \varepsilon_d$ with the restriction that other parameters are fixed, since we assume the Fermi surface keeps the form. In the case of the second parameter set, the hopping parameter changes as $t = t_{dp}^2/|\varepsilon_p - \varepsilon_d|$ so that $t$ is largely enhanced when $|\varepsilon_p - \varepsilon_d| \sim 0$. Our formulation assumes a simple case where holes enter only the lowest band and that the lowest single energy band can be expanded in powers of $1/|\varepsilon_p - \varepsilon_d|$. Therefore, our results are reliable only in the range of $|\varepsilon_p - \varepsilon_d| \geq \sim 1$ (in unit of $t_{dp}$). One may wonder if the sharp increase of $x$ with increase of $\varepsilon_p - \varepsilon_d$ in the region $|\varepsilon_p - \varepsilon_d| \geq \sim 1$ may be caused by relative increase of $V_{kk'}$ due to decrease of $t$. Some part may come this way but the dominant part is considered to come from increase of $U_{\text{eff}}$ in eq. (13)

since the simplified calculation of the $x$ curve is very well fitted with this approximate computation with $U_{\text{eff}}$.

In the case of the first parameter set with $t = t_{dp}^2/|\varepsilon_p - \varepsilon_d|$ fixed at unity, the similar limitation for validity of results should be considered as well. However, the results for the both parameter sets show qualitatively similar behaviours, i.e., SC strength parameter $x$ quickly increases as $|\varepsilon_p - \varepsilon_d|$ increases. Therefore, this behaviour is considered to give a qualitatively correct result for $x$ as a function of $|\varepsilon_p - \varepsilon_d|$ when other parameters are fixed.

## 5. SC strength parameter $x$ in the case of $\varepsilon_p - \varepsilon_d < 0$

5.1 Modified Kondo's gap equation

In the case of $\varepsilon_p - \varepsilon_d < 0$ and finite $U_p$, holes are created (annihilated) as well by $\alpha^\dagger_{-,k\sigma}$ ($\alpha_{-,k\sigma}$) in the lowest band, with which $p^\dagger_{xk\sigma}$ and $p^\dagger_{yk\sigma}$ and $U_p$ play dominant roles. This is because $u_-(k)$ diminishes quickly with decrease of $\varepsilon_p - \varepsilon_d$ in the present case, being only significant in the neighborhood of $\varepsilon_p - \varepsilon_d \sim 0$. We employ an approximation in which $d_{k\sigma} \approx u_-(k)\alpha_{-,k\sigma}$, $p_{xk\sigma} \approx v_-(k)\alpha_{-,k\sigma}$ and $p_{yk\sigma} \approx w_-(k)\alpha_{-,k\sigma}$ and get a new $V_{kk'}$. The interaction $V_{kk'}$ in this case is obtained in powers of $U_p$ and $U_d$. The first-order terms have no $d$-wave component and give no contribution to the $d$-wave SC and so we neglect them. The second-order terms are diagrammatically expressed as in Fig. 1. In difference from Kondo's case, e. g., simple Hubbard model, conversion coefficients $u_-(k)$, $v_-(k)$ or $w_-(k)$ are attached at both ends of straight lines for electrons. Summing up all terms we get

$$V_{kk'} = \frac{1}{N^2}\sum_q \frac{f_{q-k'} - f_{k+q}}{E_-(k+q) - E_-(q-k')} \left| \sum_{i=1}^{3} U(i) z_i(k) z_i(-k-q) z_i(k') z_i(q-k') \right|^2, \quad (14)$$

where $(U(1), U(2), U(3)) = (U_d, U_p, U_p)$ and $(z_1(k), z_2(k), z_3(k)) = (u_-(k), v_-(k), w_-(k))$; in eq. (14) we retain an inter-orbital interaction term which was neglected in the preceding work [5]. The lowest band is still $E_-(k)$ but its functional form is changed from that in the case of $\varepsilon_p - \varepsilon_d > 0$.

From the gap equation $\Delta_k = -\Sigma_{k'} V_{kk'} \Delta_{k'}/2E_{k'}$ which has the new $V_{kk'}$, we derive Kondo's gap equation and evaluate the SC strength $x$ in Kondo's procedure.

5.2 First band parameter set

We employ the same first band parameter set defined in Subsection 4.1 for the approximate tight-binding band for $E_-(\mathbf{k})$ as in Section 4. Figure 4 shows the results of the SC strength in the range of $\varepsilon_p-\varepsilon_d <0$ with $U_p=6$ and $U_d=8$. Values of $x$ are ~2.5 times larger than in the preceding work [5]; factor 2 comes from the correction of $V_{kk'}$ and another factor 1.25 comes from the inter-orbital term in $V_{kk'}$ which is taken into account in this work. The improvement of $x$ by a factor ~2.5 could be important depending on the situation, since it means the increase of $\Delta=\exp(-2t^2/xU_d^2)$ due to increase of $x$.

$U_d^2$ in the combination $2t^2/xU_d^2$ arised since we used a formulation in this Section parallel to that in the preceding Section. If we take $U_p/N$ as the prefactor in eq. (7), we get $\Delta=\exp(-2t^2/x'U_p^2)$ where $x'$ is the new eigenvalue in the case where we choose $U_p/N$ as the prefactor. In both cases the obtained values of $\Delta$ should be the same.

5.3 Second band parameter set

We studied the other case of second band parameter set in which $t_{dp}$ is set at 1; $t = t_{dp}^2/|\varepsilon_p-\varepsilon_d|$, $t'=0.25t$ and $t_2=-0.125t$. The ratio of $t$, $t'$ and $t_2$ is kept the same as in the case of the parameter set in Section 5.2 so that the form of the Fermi surface is similar to that of HgTl$_2$CuO$_4$ and Tl$_2$Ba$_2$CuO$_6$. As seen in Fig. 5 the $x$ value increases more quickly with decrease of $\varepsilon_p-\varepsilon_d (<0)$. This setting is considered to be more appropriate for comparison of the results with experiments, since $t_{dp}$ is observed to be roughly constant in experiments on cuprates. It clearly affirms that $x$ increases with decrease of $\varepsilon_p-\varepsilon_d (<0)$.

5.4 Comments

As seen in Figs. 4 and 5, the two cases with different band parameter sets share basic features that $x$, therefore $\Delta$ and $T_c$, increase with decrease of $\varepsilon_p-\varepsilon_d$ in the case of $\varepsilon_p-\varepsilon_d <0$ and thus assures the feature is essentially correct. As stated above the result for the second band parameter set is considered to be closer to the reality so that it assures the sharp increase of $x$ with decreasing $\varepsilon_p-\varepsilon_d$.

The results with $\varepsilon_p-\varepsilon_d<0$ indicate that the four types of SC take close values of $x$, suggesting that the most stable one may not necessarily be the $b_{1g}$ type in this model. However, since all four SC strength values increase as $|\varepsilon_p-\varepsilon_d|$ increases when $\varepsilon_p-\varepsilon_d<0$, this tendency must be correct.

In view of the change in the $\varepsilon_p-\varepsilon_d<0$ side in Fig. 4 due to the inter-orbital interaction of eq. (14) in the present version, i.e. upward shift of $x$ by 25 %, some readers may wonder how much effect is caused in the $\varepsilon_p-\varepsilon_d>0$ side in Fig. 2 if we take account of the inter-orbital interaction. In the latter case the lowest band consists almost wholly of the

*d*-orbital. Among the transformation coefficients in eqs. (5), $u_-(\boldsymbol{k})$ for the *d*-orbital is overwhelmingly larger than the two others in almost whole range of $\varepsilon_p-\varepsilon_d>0$ so that the correction due to the inter-orbital interaction should be much smaller in the $\varepsilon_p-\varepsilon_d>0$ side.

## 6. Discussion

Increase of $|\varepsilon_p-\varepsilon_d|$ in the region of $\varepsilon_p-\varepsilon_d<0$ means decrease of $\Delta V_M = e(V_M^O - V_M^{Cu})$ since $\varepsilon_p-\varepsilon_d = (A_2^O - I_3^{Cu}) + e(V_M^O - V_M^{Cu})$, where $A_2^O$ is the second electron affinity of O and $I_3^{Cu}$ is the third electron ionicity of Cu; e>0. Therefore, this relationship is consistent with the systematics reported in [1] which tell that SC strength *x* increases with decreasing $\Delta V_M$. This tendency is also in agreement with NMR/NQR results [2] that increase of $T_c$ correlates with increase of relative hole density at the planar O site. The latter should be caused by decrease of $\Delta V_M$ and so it should lead to increase of $|\varepsilon_p-\varepsilon_d|$, if $\varepsilon_p-\varepsilon_d<0$, and according to our results to increase of $\Delta$ and obviously $T_c$. If $\varepsilon_p-\varepsilon_d>0$, decrease of $\Delta V_M$ should cause decrease of $|\varepsilon_p-\varepsilon_d|$ and so $T_c$ should decrease in the opposite direction to the observation. The increasing $T_c$ with decreasing $\varepsilon_p-\varepsilon_d$ in the region of $\varepsilon_p-\varepsilon_d>0$ cannot be fitted to our results. The two correlations are consistent with our results only when $\varepsilon_p-\varepsilon_d<0$.

In the *t-J* model scheme some people may assume the positive sign for $\varepsilon_p-\varepsilon_d$. However, some researchers do not agree that the issue is fixed in this way. First of all, $\varepsilon_p$ and $\varepsilon_d$ in our model are defined in the Hamiltonian (1) and (2) without Coulomb interaction. The band splitting into upper and lower Hubbard bands is not taken into account in $\varepsilon_p$ and $\varepsilon_d$ here. For evaluation of $\varepsilon_p$ and $\varepsilon_d$ we have to rely on the optical experiments, such as EELS, together with strong *U* theories of atomic spectra [11]. Some researchers have estimated $\varepsilon_p$ and $\varepsilon_d$ in this way and suggest $\varepsilon_p-\varepsilon_d$ takes a small negative value. Therefore, our scheme with the negative $\varepsilon_p-\varepsilon_d$ has a possibility to live on.

With increase of the distance of the apex oxygen away from the $CuO_2$ plane, cuprates are known to increase $T_c$ [12]. This is parallel with our model's results if $\varepsilon_p-\varepsilon_d<0$. The accompanying raise of $\varepsilon_d$, due to going away of the anion, and according decrease of $\varepsilon_p-\varepsilon_d$, i. e., increase of $|\varepsilon_p-\varepsilon_d|$ should tend to increase $T_c$. At least this can be a positive factor for increasing $T_c$ with increasing apex oxygen height.

When the Hubbard model is applied to cuprates, $T_c$ is observed to be enhanced quickly with increase of *U* [13]. In the modified Hubbard model derived by us it is possible to define an effective *U* and ascribe the increase of $T_c$ to the increase of the effective $U$, then increase of $U$ to the increase of $|\varepsilon_p-\varepsilon_d|$. This argument is quantitatively successful in the case of $\varepsilon_p-\varepsilon_d>0$. It works also in the case of $\varepsilon_p-\varepsilon_d<0$, if we separately compute the contributions from $v_-(\boldsymbol{k})$-related terms and $w_-(\boldsymbol{k})$-related terms.

The above-mentioned accordances indicate the plausibility to explain the high-$T_c$ by the increase of $|\varepsilon_p-\varepsilon_d|$ in the region of of $\varepsilon_p-\varepsilon_d<0$ in such cuprate superconductors as $HgTl_2CuO_4$ and $Tl_2Ba_2CuO_6$ with the assumption that our results can be extrapolated to the intermediate coupling case at least qualitatively.

Sakakibara et al. proposed another scheme for the increase of $T_c$ with the increasing apex distance [14]. They focus the role of the $d_{z^2}$ level lying lower than the $d_{x^2-y^2}$ level in the electron picture; the hole occupation in the $d_{z^2}$ band is asserted to decrease $T_c$. The $O2p_o$ level does not appear explicitly in this scheme. It is necessary to clarify how effectively each theory works in related superconductors.

Three-layer cuprates have a tendency to have higher $T_c$ in the inner layer than in the outer layers [15]. Together with their pressure dependence, their $T_c$'s are more naturally explained as due to the variation of $\Delta V_M$ between the inner plane copper and oxygen sites. In view of such an example, each systematics deserves to be examined carefully case by case.

On the basis of the ionic model and the Madelung potential computation, Ohta et al. [16] examined the sytematics of $T_c$ of high-$T_c$ cuprates. They found that $\Delta V_A$, which is the difference between the Madelung potential at the apex oxygen and that at the neighboring planar oxygen, increases in a clear-cut correlation with the highest $T_c$ among each kind of cuprates. They ascribed this increase of $T_c$ to stabilization of the Zhang-Rice singlet in the strong-coupling framework of the theory in a qualitative success. However, we consider that they have still much work to be done for supporting their hypothesis microscopically.

Since the theory of high-$T_c$ SC is not yet established, theories of the types, other than strong-coupling, such as the present moderate-coupling one, can be still valuable, if they describe some of systematics successfully.

## 7. Summary

Relationships between $T_c$ and the $2p_\sigma$-orbital level of planar oxygen in cuprate high-$T_c$ superconductors are clearly shown by Torrance et al. and also by Zheng et al. In the present work we have studied the SC transition of the $d$-$p$ model which explicitly takes account of the $2p_\sigma$-orbital level $\varepsilon_p$, focusing on the relation between the SC strength and $\varepsilon_p$. Employing the hole picture and treating the case in which ~1.15 holes/unit-cell occupy the lowest band of the one-hole Hamiltonian, we derived the tight-binding band for the hole and the effective Coulomb interaction between holes. In this Hubbard model in a wide sense the Coulomb interaction is non-local, or has a rather strong $k$-dependence, and depends on $\varepsilon_p$-$\varepsilon_d$, where $\varepsilon_d$ is the Cu$3d_{x^2-y^2}$ orbital level.

Modifying Kondo's gap equation of the weak-coupling Hubbard model for the present broad-sense Hubbard model, we computed the so-call SC strength $x$ with the band parameters of the HgTl$_2$CuO$_4$ and Tl$_2$Ba$_2$CuO$_6$ cuprates with on-site Coulomb energies $U_d$ and $U_p$ chosen by hand. $x$ increases with increase of $|\varepsilon_p$-$\varepsilon_d|$, indicating strengthening SC. We found that the present result qualitatively explains the above-mentioned two systematics if we choose $\varepsilon_p$-$\varepsilon_d<0$. This suggests that $\varepsilon_p$-$\varepsilon_d<0$ in high-$T_c$ cuprates such as HgTl$_2$CuO$_4$ and Tl$_2$Ba$_2$CuO$_6$. In the case of $\varepsilon_p$-$\varepsilon_d<0$, the four even-parity $d$-wave SC states are obtained which have roughly equal stability.

**Acknowledgements**

The authors are grateful to Professor J. Kondo for kind discussions. They deeply thank Professor S. Tajima for providing us helpful vital information on the values of $\varepsilon_p$ and $\varepsilon_d$ of cuprates. They are also thankful to Professor T. Tohyama for useful discussions.

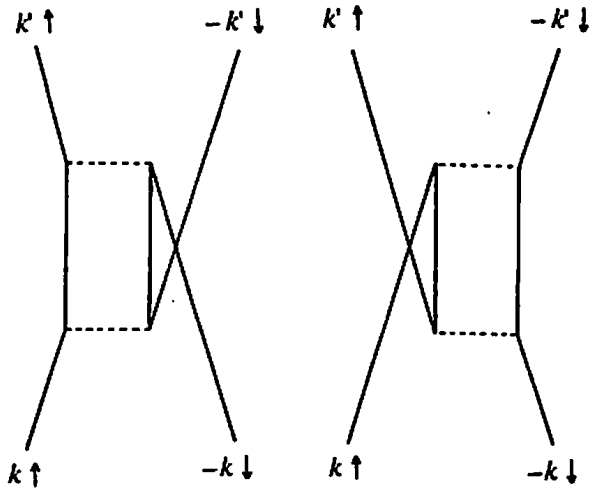

Fig. 1. Diagrams contributing to the pairing interaction in the second order in the Coulomb interaction (dashed lines). From Fig. 1 in ref. [5].

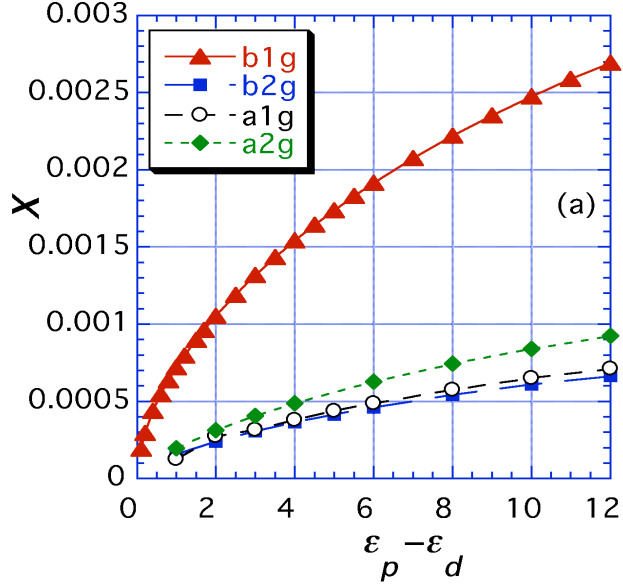

Fig.2. SC strength parameter $x$ vs $\varepsilon_p-\varepsilon_d$ in its positive side of $\varepsilon_p-\varepsilon_d$. $U_d$=8 and $U_p$=6 are employed. Band parameters are $t$=1, $t'$=0.25 and $t_2$=−0.125. About the energy unit see the text. Results in the large-value range of $|\varepsilon_p-\varepsilon_d|$ is reliable, but the result in the range of $|\varepsilon_p-\varepsilon_d|<1$ should not be taken as it is, since $E_-(\bm{k})$ may cross other higher bands which is not assumed in the theory. Leading $\theta$-dependences of the four types of $d$-wave gap parameters are $b_{1g}$: $\cos2\theta$, $b_{2g}$: $\sin2\theta$, $a_{1g}$: $\cos4\theta$, $a_{2g}$: $\sin4\theta$, respectively.

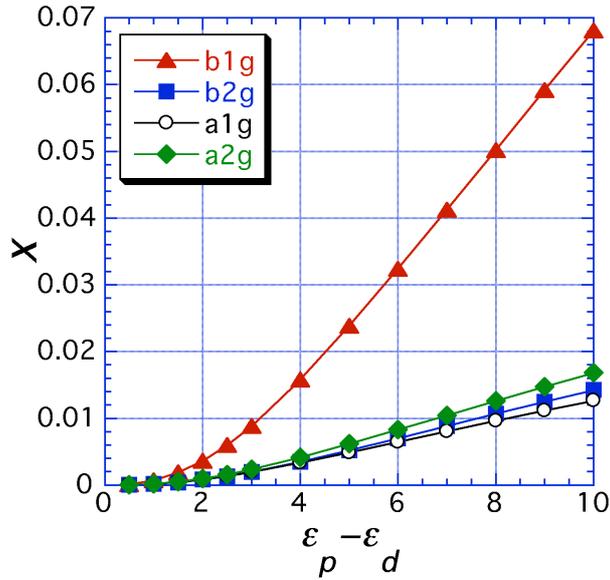

Fig. 3. SC strength parameter $x$ vs $\varepsilon_p-\varepsilon_d$ in its positive side of $\varepsilon_p-\varepsilon_d$. $U_d$=8 and $U_p$=6 are employed. Band parameters are $t_{dp}$=1, $t'$=0.25$t$ and $t_2$=−0.125$t$;   $t=t_{dp}^2/(\varepsilon_p-\varepsilon_d)$.

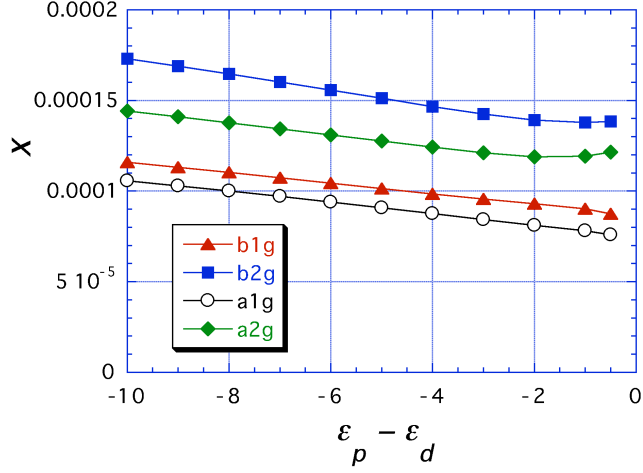

Fig. 4. SC strength parameter $x$ vs $\varepsilon_p - \varepsilon_d$ in its negative side of $\varepsilon_p - \varepsilon_d$. Employed $U_d=8$ and $U_p=6$ are employed. Band parameters are $t=1$, $t'=0.25$ and $t_2=-0.125$; $t=t_{dp}^2/|\varepsilon_p - \varepsilon_d|$. For the unit see the text. Leading $\theta$-dependences of the four types are $b_{1g}$: $\cos 2\theta$, $b_{2g}$: $\sin 2\theta$, $a_{1g}$: $\cos 4\theta$, $a_{2g}$: $\sin 4\theta$, respectively.

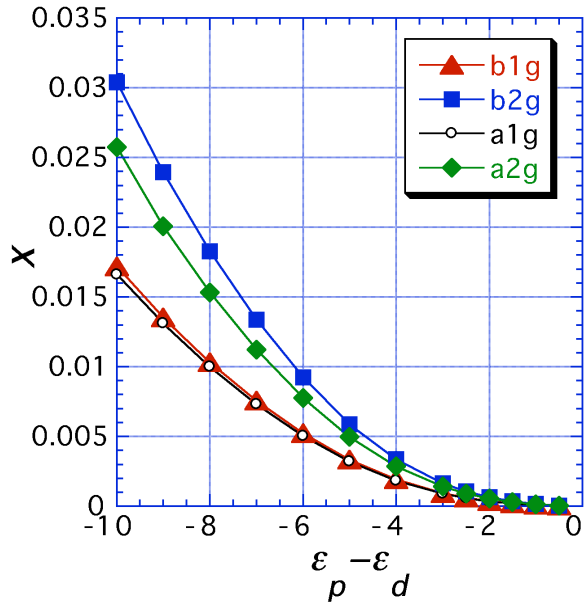

Fig.5. SC strength parameter $x$ vs $\varepsilon_p - \varepsilon_d$ in its negatve side of $\varepsilon_p - \varepsilon_d$. $U_d=8$ and $U_p=6$ are employed. Band parameters are $t_{dp}=1$, $t'=0.25t$ and $t_2=-0.125t$; $t=t_{dp}^2/|\varepsilon_p - \varepsilon_d|$.